\RequirePackage[2020-02-02]{latexrelease}
\documentclass[preprint,prd,amsmath,amssymb,floatfix,nofootinbib]{revtex4}

\usepackage[colorlinks=true,urlcolor=black]{hyperref}

\hypersetup{backref = true,pagebackref = true,hyperindex = true, colorlinks = true,
  breaklinks = true, urlcolor = blue, linkcolor = blue, bookmarks = true,
  bookmarksopen = true, citecolor=red}

\usepackage{bm}
\usepackage{color}
\usepackage{dcolumn}
\usepackage{graphics}
\usepackage{graphicx}
\usepackage{subfigure}
\usepackage{slashed}
\usepackage{placeins}
\usepackage{pdfpages}
\usepackage{eso-pic}
\usepackage{everyshi}
\usepackage{pdflscape}
\allowdisplaybreaks[4]
\usepackage{xcolor}

\begin{document}
\setlength{\unitlength}{1mm}
\newcommand{\z}{&&\hspace*{-1cm}}
\newcommand{\zz}{&&\hspace*{-4cm}}
\newcommand{\ep}{\varepsilon}
\newcommand{\cita} [1] {$^{\hbox{\scriptsize \cite{#1}}}$}
\newcommand{\prepr}[1] {\begin{flushright} {\bf #1} \end{flushright} \vskip 1.5cm}

\newcommand{\bea}{\begin{eqnarray}}
\newcommand{\eea}{\end{eqnarray}}
\newcommand{\be}{\begin{equation}}
\newcommand{\ee}{\end{equation}}
\newcommand{\MSbar}{\overline{\rm MS}}
\newcommand{\as}{\alpha_s}
\newcommand{\asMZ}{\alpha_s(M^2_Z)}
\newcommand{\dd}{\mathrm{d}}
\newcommand{\ar}{a_s}
\newcommand{\oar}{\overline{a}_s}

\title{Fractional Analytic QCD  beyond Leading Order  in timelike region}
       \author{A.\ V.~Kotikov$^{1}$ and I.A.~Zemlyakov$^{1,2}$
       }
       \affiliation{
$^1$Bogoliubov Laboratory of Theoretical Physics, Joint Institute for Nuclear Research, 141980 Dubna, Russia.\\
         $^2$Dubna State University,
Dubna, Moscow Region, Russia }

\date{\today}

\begin{abstract} 

   In this paper we show that, as in the spacelike case, the inverse logarithmic
expansion is applicable for all values of the argument of the analytic coupling constant.
   We present two different approaches, one of which is based primarily on trigonometric
   functions, and the latter is based on dispersion integrals.
   The results obtained up to the 5th order of perturbation theory, have a compact form and their acquiring is much easier than the methods that have been used before.
   As an example, we
   apply our results to study the Higgs boson decay into a $b\overline{b}$ pair.
  
\end{abstract}

\maketitle

\section{ Introduction }
\label{Intro}

  The perturbative expansion in QCD works well only for estimation of the quantities in the region of large squared momentum $Q^2$ (here and further
  $Q^2=-q^2$, where $q^2$ -- transfered momentum in the Euclidean domain for space-like processes). However, for transfered momenta less than 1 GeV$^2$, the situation
  changes dramatically. The reason for this is the presence of the singularity of  the coupling constant (couplant) $\alpha_{\rm s}(Q^2)$ at the point $Q^2=\Lambda^2$
  which is widely known as Landau (ghost) pole.
  This singularity is especially important when we expand various physical observables in terms of the couplant, which makes
    the behavior of the observables
    non-analytic in the $Q^2$-plane.
   For the correct description of QCD observables in the region of small $Q^2$ values, it is necessary to construct a new everywhere continuous couplant.
  
  The renormalization group (RG) method allows to sufficiently improve the expressions obtained in the frame of perturbation theory (PT).
 To show   that the RG method cannot solve the abovementioned problem, we first write differential equation
   \be
   \frac{d}{dL} \, \overline{a}_s(Q^2) = \beta(\overline{a}_s),~~\overline{a}_s(Q^2)=\frac{\alpha_s(Q^2)}{4\pi},~~ L=\frac{Q^2}{\Lambda^2},~~ 
\label{oas}
\ee
 with the QCD $\beta$-function
\be
\beta(\overline{a}_s) ~=~ -\sum_{i=0} \beta_i \overline{a}_s^{i+2} 
=-\beta_0  \overline{a}_s^{2} \, \Bigl(1+\sum_{i=1} b_i \beta_0^i \overline{a}_s^i\Bigr),~~ b_i=\frac{\beta_i}{\beta_0^{i+1}}\, ,
\label{beta}
\ee
where the first fifth coefficients, i.e. $\beta_i$ with $i\leq 4$, are exactly known \cite{Baikov:2016tgj,Herzog:2017ohr,Luthe:2017ttg}.
Here we use the following
definition of strong couplant:
\be
\ar(Q^2)=\frac{\beta_0\alpha_s(Q^2)}{4\pi}\,=\beta_0\,\overline{a}_s(Q^2)\,, 
\label{as}
\ee
where we absorb
the first coefficient of the  QCD $\beta$-function into the $\ar$ definition, as is usually
the case of  
analytic couplants (see, e.g., Refs. \cite{ShS}-\cite{Cvetic:2008bn}).

Solving Eq.(\ref{oas}) for $\overline{a}_s(Q^2)$
with the only leading order (LO) term on the right side, one can obtain the one-loop expression
\be
\ar^{(1)}(Q^2) = \frac{1}{L}\, ,
\label{asLO}
\ee
i.e. $\ar^{(1)}(Q^2)$ contains the pole at $Q^2=\Lambda^2$
  that indicates the inability of the RG approach to remove the Landau pole.

In the timelike region  ($q^2 >0$) (i.e., in the Minkowski space), the determination of the running coupling turns out to be quite difficult.
The reason for the problem is that, strictly speaking, the expansion of perturbation theory in QCD cannot be determined directly in this area.
Indeed, since the early days of QCD, much effort has been made to determine the appropriate coupling parameter in the  Minkowski space to describe
important timelike processes such as, for example, the $e^+e^-$-annihilation into hadrons, quarkonium and $\tau$-lepton decays into hadrons. Most of the
attempts (see, for example, \cite{Pennington:1981cw}) were based on the analytical continuation of the strong couplant from the deep Euclidean region,
where QCD perturbative calculations can be performed, to the Minkowski space, where physical measurements are performed.
Over the time, it became clear that in the infrared (IR) regime,
the strong couplant can reach a stable fixed point and stop increasing. This behavior would
imply that the color forces can saturate at low momenta. So, for example, Cornwall \cite{Cornwall:1981zr} already in 1982 obtained the appearance of
the gluon effective mass, which behaves as IR regulator in the region of small momenta. Similar results were obtained by others in subsequent 
years (see, for example, \cite{Parisi:1979se}) using different methods.

In other developments, analytical expressions for LO couplant directly in the Minkowski space were obtained \cite{Krasnikov:1982fx}
using an integral transformation from the spacelike to the timelike region for
the Adler D-function (more information can be found in Ref.
\cite{Bakulev:2000uh}).

The systematic approach, called the analytical perturbation theory (APT), arose in the Shirkov and Solovtsov  studies \cite{ShS}.
In this paper authors proposed to use new everywhere continuous analytic couplant  $A_{\rm MA}(Q^2)$ in the form of spectral integral
    \be
A^{(i)}_{\rm MA}(Q^2) 
= \frac{1}{\pi} \int_{0}^{+\infty} \, 
\frac{ d \sigma }{( \sigma + Q^2)} \, r^{(1)}_{\rm pt}( \sigma) \, ,
\label{disp_MA_LO}
\ee
which is directly related with the appropriate PT order
via the spectral function $r_{\rm pt}(s)$ 
\be
r^{(i)}_{\rm pt}( \sigma)= {\rm Im} \; a_s^{(i)}(-\sigma - i \epsilon) \,.
\label{SpeFun_LO}
\ee
Similarly, the analytical images of a running coupling in Minkowski space are defined using another linear operation
    \be
U^{(i)}_{\rm MA}(s) 
= \frac{1}{\pi} \int_{s}^{+\infty} \, 
\frac{ d\sigma }{\sigma} \, r^{(i)}_{\rm pt}( \sigma) \, ,
\label{disp_MAt_LO}
\ee
This method, called as {\it Minimal Approach} (MA) (see, e.g., \cite{Cvetic:2008bn}), contains spectral function of pure perturbative nature.
\footnote{An overview of other similar approaches can be found in \cite{Bakulev:2008td} including approaches \cite{Nesterenko:2003xb,Nesterenko:2004tg} close to APT.}

The Analytic couplants $A_{\rm MA}(Q^2)$ and $U_{\rm MA}(s)$ take almost the same values as $a_s(Q^2)$ when $Q^2(s)>>\Lambda^2$ and completely different finite values at
$Q^2 \leq \Lambda^2$.
Moreover, the MA couplants $A_{\rm MA}(Q^2)$ and $U_{\rm MA}(Q^2)$ are related each other as
  \cite{Bakulev:2006ex}
\be
A^{(i)}_{\rm MA}(Q^2) 
= \int_{0}^{+\infty} \, 
\frac{ d\sigma \, Q^2}{(\sigma + Q^2)^2} \, U^{(i)}_{\rm MA}( \sigma),~~ U^{(i)}_{\rm MA}( s) =\frac{1}{2\pi i} \,  \int_{-s-i\ep}^{-s+i\ep} \, 
\frac{ d\sigma }{\sigma} \, A^{(i)}_{\rm MA}( \sigma) \, .
\label{disp_MApt_LO}
\ee
The APT were extended for the case of non-integer power of couplant, which appears in QFT framework for quantities
with  non-zero anomalous dimensions (see the famous papers \cite{BMS1,Bakulev:2006ex,Bakulev:2010gm},
some previous study \cite{Karanikas:2001cs} and
reviews in Ref. \cite{Bakulev:2008td}). For these purposes the fractional APT (FAPT) was developed. Due to the complexity of FAPT, the main results here until recently were
obtained mostly in LO, however, it was also used in higher orders by re-expanding the corresponding coupling constants in the
terms of LO ones, as well as using some approximations.

Following our recent paper  \cite{Kotikov:2022sos} devoted to the couplant in the Euclidean domain, in this article
  we extend the FAPT in the Minkowski space to higher PT orders
  using the so-called
$1/L$-expansion of the usual couplant. For an ordinary couplant,
this expansion is valid only for the
large values of  $L$, i.e. for $Q^2>>\Lambda^2$; however, as it was shown in \cite{Kotikov:2022sos},
if we consider an analytic couplant, this
expansion is applicable throughout whole axis of squared transfered momentum.  This becomes possible due to the smallness of the
corrections for MA couplant which disappear when $Q^2 \to \infty$ and also $Q^2 \to 0$.
\footnote{The absence of high-order corrections for $Q^2 \to 0$ was also discussed in Refs. \cite{ShS,MSS,Sh}.} Thus, only in the region $Q^2 \sim \Lambda^2$ corrections turn out to be important enough
(see also detailed discussions in Section 3 below). 

Below we represent two different forms for the MA couplant
in the Minkowski space
calculated up to the 5th order of PT both of which
 contain the coefficients of the QCD $\beta$-function as parameters
(some short version with the results based on  the first three orders can be found in Ref. \cite{Kotikov:2022vnx}).

 The paper is organized as follows. In Section \ref{strong} we shortly review the basic properties of the usual strong couplant,
 its fractional derivatives (i.e. the $\nu$-derivatives) and the
$1/L$-expansions, which
can be represented as some operators acting on the $\nu$-derivatives of the LO strong couplant. This was the key idea of the paper \cite{Kotikov:2022sos}, which makes it possible
here to construct $1/L$-expansions of the $\nu$-derivatives of MA couplant in the Minkowski space for high-order perturbation theory, see Section \ref{MA}. In  Section \ref{IntegralRe}
we applied our new derivative operators to
  an integral representations of the MA couplant in the Minkowski space and in this manner continued it at the high PT orders.
Section \ref{HbbDecay} contains an application of this approach to the the Higgs boson decay into a $b\overline{b}$ pair.
In conclusion, some final discussions are given. In addition, 
we have several Appendices.
Appendix A  presents some alternative results for the $\nu$-derivatives of the MA couplant $U^{(1)}_{\rm MA}(s)$, which may be
useful for some applications. Some details related with the derivation of the coefficients of the running quark mass are gathered in Appendix B.
Appendix C contains formulas for restoring non-integer  $\nu$-powers of the
usual strong couplant as series of its  $(\nu+m)$-derivatives.

\section{Strong coupling constant and it fractional derivatives}
\label{strong}

The strong couplant $a_s(Q^2)$ can be represented as $1/L$-series when $Q^2>>\Lambda^2$. Here we give
the first five terms of the expansion in an agreement with the number of known coefficients $\beta_i$ in the following short form 
\be
  a^{(1)}_{s}(Q^2) = \frac{1}{L},~~
a^{(i+1)}_{s}(Q^2) = 
a^{(1)}_{s}(Q^2) + \sum_{m=2}^{i+1} \, \delta^{(m)}_{s}(Q^2)
\,,~~(i=0,1,2,...)
\label{as}
\ee
where  L is defined in  Eq. (\ref{oas}).

The corresponding corrections $\delta^{(m)}_{s}(Q^2)$ are represented in \cite{Kotikov:2022sos}.
  At any PT order,
  the couplant $\ar(Q^2)$ contains its own parameter $\Lambda$
of dimensional transmutation, which is fitted from experimental data for every single case.
 
The coefficients $\beta_i$ depend on the number $f$ of flavors, which increases or decreases at thresholds $Q^2_f \sim m^2_f$, where some new quark appears at
$Q^2 > Q^2_f$. Here $m_f$ is the $\overline{MS}$ mass of $f$ quark, for example, $m_b=4.18 +0.003-0.002$ GeV  from PDG20 \cite{PDG20}.
\footnote{Strictly speaking, the quark masses are $Q^2$-dependent  in $\overline{MS}$-scheme and $m_f=m_f(Q^2=m_f^2)$. However, the $Q^2$-dependence is quite slow and it is
  not shown in the present study.}
Thus, the couplant
  $a_s$ is $f$-dependent and its $f$-dependence can be incorporated into $\Lambda$, as $\Lambda^f$, where f indicates the number of active flavors.
In the $\overline{MS}$ scheme, the relations between $\Lambda_{i}^{f}$ and $\Lambda_{i}^{f-1}$
  are known up to the four-loop order
\cite{Chetyrkin:2005ia,Schroder:2005hy,Kniehl:2006bg} and they are usually used
at $Q_f^2=m_f^2$, where the relations are simplified (for a recent review, see e.g. \cite{FLAG,Enterria}).

Below we mainly deal with the region of low $Q^2$, where the only 3 first lightest quarks appear. Since in this case we will
  use the set of $\Lambda_{i}^{f=3}$ $(i=0,1,2,3)$ taken from the recent Ref. \cite{Chen:2021tjz}.
Further, since we will
consider the $H \to b\overline{b}$ decay as an application, 
we will use also the results for $\Lambda_{i}^{f=5}$
taken also from \cite{Chen:2021tjz}
\bea
&&\Lambda_0^{f=3}=142~~ \mbox{MeV},~~\Lambda_1^{f=3}=367~~ \mbox{MeV},~~\Lambda_2^{f=3}=324~~ \mbox{MeV},~~\Lambda_3^{f=3}=328~~ \mbox{MeV}\,, \nonumber \\
&&\Lambda_0^{f=5}=87~~ \mbox{MeV},~~\Lambda_1^{f=5}=224~~ \mbox{MeV},~~\Lambda_2^{f=5}=207~~ \mbox{MeV},~~\Lambda_3^{f=5}=207~~ \mbox{MeV}\,.
\label{Lambdas}
\eea
We use also $\Lambda_4=\Lambda_3$, since in the highest orders $\Lambda_i$ values become very similar.

\subsection{Fractional derivatives}
\label{FracDe}

As it was done in \cite{Cvetic:2006mk,Cvetic:2006gc},
  we firstly introduce the derivatives of couplant (in the $(i+1)$-order of PT)
\be
\tilde{a}^{(i+1)}_{n+1}(Q^2)=\frac{(-1)^n}{
  n!} \, \frac{d^n a^{(i+1)}_s(Q^2)}{(dL)^n} \, ,
\label{tan+1}
\ee
which is a key element
in construction of FAPT
(see e.g. Ref. \cite{Kotikov:2022swl} and discussions therein).

The derivatives $\tilde{a}_{n}(Q^2)$ can be  successfully used instead of $\ar$-powers in the decomposition of QCD observables. Although every
derivative decreases the power of $\ar$, but it produces  the additional $\beta$-function $\sim \ar^2$, appeared from the term $d\ar/dL$.
At LO,
the series of derivatives exactly coincide with the series of powers. Beyond LO, 
the relation between $\tilde{a}_{n}(Q^2)$ and $\ar^{n}(Q^2)$ was established \cite{Cvetic:2006gc,Cvetic:2010di}
(the corresponding expansion $a^{n+1}(Q^2)$ in the terms $\tilde{a}^{(i+1)}_{n+m+1}(Q^2)$
can be found in Appendix C)
and extended to the fractional case,
where $n $ is replaced for a non-integer $\nu $, in Ref. \cite{GCAK}.
The results for evaluation of
  $\tilde{a}^{(i+1)}_{n+1}(Q^2)$ are shown in (\ref{tan+1})
was considered in details in Appendix B of \cite{Kotikov:2022sos}.

Here we write only the final results of calculations,
which are represented in the form similar to given in Eq.(\ref{as})
\footnote{The expansion (\ref{tdmp1N}) is very similar to those used in Refs. \cite{BMS1,Bakulev:2006ex} for the expansion of
  ${\bigl({a}^{(i+1)} _{s}(Q^2)\bigr)}^ {\nu}$ in terms of powers of $a^{(1)}_{s}(Q^2)$.}
:
\bea
&&\tilde{a}^{(1)}_{\nu}(Q^2)={\bigl(a^{(1)}_{s}(Q^2)\bigr)}^{\nu} = \frac{1}{L^{\nu}},~~
\tilde{a}^{(i+1)}_{\nu}(Q^2)=\tilde{a}^{(1)}_{\nu}(Q^2) + \sum_{m=1}^{i}\, C_m^{\nu+m}\, \tilde{\delta}^{(m+1)}_{\nu}(Q^2),~~\nonumber\\
&&\tilde{\delta}^{(m+1)}_{\nu}(Q^2)=
\hat{R}_m \, \frac{1}{L^{\nu+m}},~~C_m^{\nu+m}=\frac{\Gamma(\nu+m)}{m!\Gamma(\nu)}\,,
\label{tdmp1N}
\eea
where
\bea
&&\hat{R}_1=b_1 \Bigl[\hat{Z}_1(\nu)+ \frac{d}{d\nu}\Bigr],~~
\hat{R}_2=b_2 + b_1^2 \Bigl[\frac{d^2}{(d\nu)^2} +2 \hat{Z}_1(\nu+1)\frac{d}{d\nu} + \hat{Z}_2(\nu+1 )\Bigr], \nonumber\\
&&\hat{R}_3=\frac{b_3}{2} + 3b_2b_1\Bigl[Z_1(\nu+2)-\frac{11}{6}+\frac{d}{d\nu}\Bigr]\nonumber \\
&&+ b_1^3 \Bigl[ \frac{d^3}{(d\nu)^3}+3\hat{Z}_1(\nu+2) \frac{d^2}{(d\nu)^2} +3 \hat{Z}_2(\nu+2)\frac{d}{d\nu} + \hat{Z}_3(\nu+2 )\Bigr], \nonumber\\
&&\hat{R}_4=\frac{1}{3}\,\bigl(b_4+5b_2^2\bigr) + 2b_3b_1\Bigl[Z_1(\nu+3)-\frac{13}{6}+\frac{d}{d\nu}\Bigr]\nonumber \\
&&+ 6b_1^2b_2 \Bigl[ \frac{d^2}{(d\nu)^2}+2\left(Z_1(\nu+3)-\frac{11}{6}\right) \frac{d}{d\nu} + Z_2(\nu+3) -\frac{11}{3} \, Z_1(\nu+3)+\frac{38}{9}
    \Bigr]\nonumber\\
&&+ b_1^4 \Bigl[\frac{d^4}{(d\nu)^4}+4\hat{Z}_1(\nu+3) \frac{d^3}{(d\nu)^3} +6\hat{Z}_2(\nu+3) \frac{d^2}{(d\nu)^2}
  +4 \hat{Z}_3(\nu+3)\frac{d}{d\nu} + \hat{Z}_4(\nu+3 )\Bigr]\, .
\label{hR_i}
\eea

  The representation (\ref{tdmp1N}) of the $\tilde{\delta}^{(m+1)}_{\nu}(Q^2)$ corrections as $\hat{R} _m$-operators plays a very important role in this paper.
\footnote{The results for $\hat{R}_m$-operators contain the transcendental principle \cite{Kotikov:2000pm}: the corresponding functions $\hat{Z}_k(\nu)$ ($k \leq m$)
  contain the Polygamma-functions $\Psi_k(\nu)$ and their products, such as $\Psi_{k-l}(\nu)\Psi_l(\nu)$, and also with a larger number of factors)
  with the same total index $k$. However, the importance of this property is not clear yet.}
Hereinafter, acting these operators on the analytic couplant in the Minkowski space, we will obtain the results for high-order corrections.

\section{Minimal analytic coupling in Minkowski space}
\label{MA}

There are several ways to obtain analytical versions of the strong couplant $a_s$ (see, e.g. \cite{Bakulev:2008td}). Here we will follow
the MA approach \cite{ShS,MSS,Sh} as discussed in Introduction.
To the fractional case,  the MA approach was generalized by Bakulev, Mikhailov and Stefanis (hereinafter referred to as the BMS approach), that is
presented
in three famous papers \cite{BMS1,Bakulev:2006ex,Bakulev:2010gm}
(see also the previous paper \cite{Karanikas:2001cs}, the reviews \cite{Bakulev:2008td,Cvetic:2008bn} and Mathematica package in \cite{Bakulev:2012sm}). 

We first show the leading order BMS results, and later we will go beyond LO, following our results for the usual strong couplant obtained in the previous section
(see Eq. (\ref{tdmp1N})).

\subsection{LO}
\label{LO}

The LO MA coupling $U^{(1)}_{{\rm MA},\nu}(s)$ in the Minkowski space
has the following form \cite{Bakulev:2006ex}
\be
U^{(1)}_\nu({\rm s})=\tilde{U}^{(1)}_\nu({\rm s})=\frac{\sin[(\nu-1)\,g(s)]
}{\pi(\nu-1)(\pi^2+ L^2_s)^{(\nu-1)/2}}
  \, ,\, (\nu > 0)\, ,
\label{mainexpr}
\ee
where
\be
L_s=\ln\dfrac{s}{\Lambda^2},~~g(s)= \arccos\left(\frac{L_s}{\sqrt{\pi^2+ L^2_s}}\right) \, .
\label{Ls}
\ee
The fact that Eq.(\ref{mainexpr}) is applicable only for $\nu>0$ will be discussed later. 

\begin{figure}[h!]
	\centering
	\includegraphics[width=0.68\textwidth]{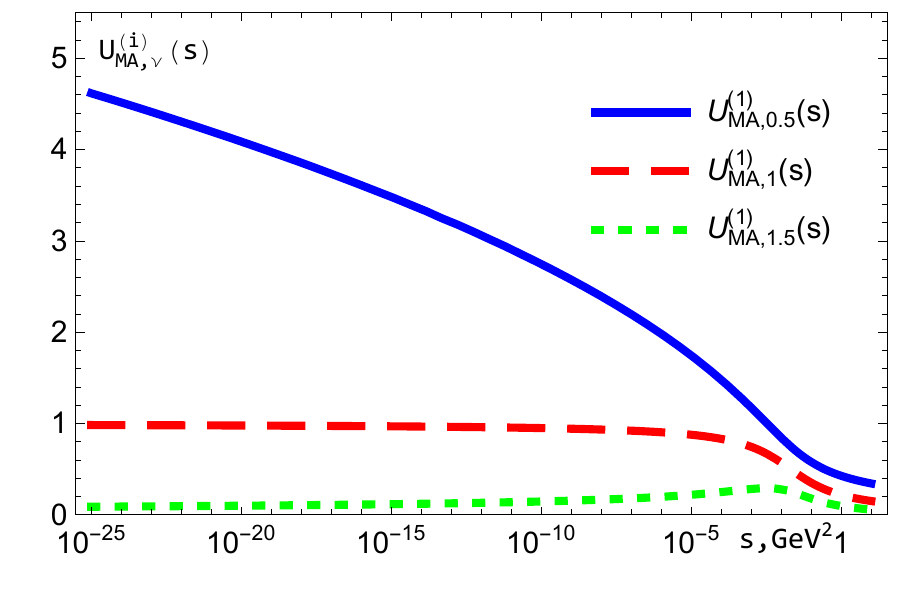}
	\caption{The results for $U^{(1)}_{\rm MA,\nu}(Q^2)$ with $\nu=0.5,1,1.5$ in logarithmic scale.}
	\label{fig:logunu}
\end{figure}

For the cases $\nu=0.5,1,1.5$, $U^{(1)}_{{\rm MA},\nu}(Q^2)$ is shown in Fig. \ref{fig:logu135}.
  Strictly speaking, the value of the parameter $\Lambda$ is obtained by fitting experimental data.
  To obtain its values
  (one of the two MA couplants $A_{\rm MA}(Q^2)$ and $U_{\rm MA}(s)$ can be fitted as they are very close to each other,
  as will be shown on Figs. \ref{fig:comparauN} and \ref{fig:aurel} below)
  within the framework of analytical QCD, it is necessary to fit experimental data
  for various processes
  \footnote{One of the most important applications is  fitting 
experimental data for the DIS structure functions (SFs) $F_2(x,Q^2)$ and $F_3(x,Q^2)$ (see, e.g., Refs.
\cite{Kotikov:2010bm,PKK,Kotikov:2015zda,KK2001} and
\cite{KKPS1,KPS}, respectively).
One can use the $\nu$-derivatives of the MA couplant $\tilde{A}_{\rm MA,\nu}^{(i)}(Q^2)$,
which is indeed possible, because when fitting
we study the SF Mellin moments (following Ref.
\cite{Barker}) and only at the end reconstruct the SF themselves.
This differs from the more popular approaches \cite{NNLOfits} based on numerical solutions of
the Dokshitzer-Gribov-Lipatov-Altarelli-Parisi (DGLAP) equations \cite{DGLAP}.
In the case of using the \cite{Barker}  approach, the $Q^2$-dependence of the SF moments is known exactly  in analytical form (see, e.g., \cite{Buras}): it
can be expressed in terms of the $\nu$-derivatives
$\tilde{A}_{\rm MA,\nu}^{(i)}(Q^2)$, where the corresponding $\nu$-variable becomes to be $N$-dependent (here $N$ is the Mellin moment number),
and the using of the $\nu$-derivatives should be crucial.
Beyond LO, in order to obtain complete analytic results for Mellin moments, we will use their analytic continuation \cite{KaKo}.}
by using,
    for instance, formulas obtained in this paper that simplify the form of higher-order terms. This, however, requires additional special
    research. In this article we use the values $\Lambda_{f=3}$ and $\Lambda_{f=5}$ (see Eq. (\ref{Lambdas}))
     obtained in the framework of a conventional perturbative QCD since PT and FAPT couplants must coincide in the limit of large $Q^2$ and this requirement is fulfilled.
%
It is clearly seen that at low $Q^2$ $U^{(1)}_{{\rm MA},\nu}(Q^2)$ agrees with its asymptotic values:
\be
U^{(1)}_{{\rm MA},\nu}(Q^2= 0) = \left\{
\begin{array}{c}
0 ~~\mbox{when}~~ \nu >1, \\
1 ~~\mbox{when}~~ \nu =1, \\
\infty ~~\mbox{when}~~ \nu <1, 
\end{array}
\right.
   \label{AQ=0}
\ee
obtained in Ref. \cite{Ayala:2018ifo}.
The corresponding results in the Euclidean space for $A^{(1)}_{{\rm MA},\nu}(Q^2)$ $\nu=0.5,1,1.5$ were numerically obtained and shown on Fig.1 in \cite{Kotikov:2022sos}.
They are very close to those shown above in
Fig. \ref{fig:logunu}.
  Moreover, the asymptotic values
  of $A^{(1)}_{{\rm MA},\nu}(Q^2= 0)$ and $U^{(1)}_{{\rm MA},\nu}(s= 0)$ are completely identical to each other.

\subsection{Beyond LO}
\label{BELO}

Hereafter we repeat for $U_{{\rm MA},\nu}(s)$ the procedure that was applied to $\tilde{a}_{\nu}^{(i)}(Q^2)$.
For this purpose,
following to the
representation (\ref{mainexpr}) for the LO MA couplant in the Minkowski space, 
we consider
its derivatives
\be
\tilde{U}_{{\rm MA},n+1}(s)=\frac{(-1)^n}{
  n!} \, \frac{d^n U_{\rm MA}(s)}{(dL_s)^n} \, .
\label{tanMA+1}
\ee

Using the results (\ref{tdmp1N}) for the usual couplant we have
\bea
&&\tilde{U}^{{\rm MA},(i+1)}_{\nu}({\rm s})=\tilde{U}^{(1)}_{{\rm MA},\nu}({\rm s}) + \sum_{m=1}^{i}\, C_m^{\nu+m}\, \tilde{\delta}^{(m+1)}_{\nu}({\rm s}),~~\nonumber\\
&&\tilde{\delta}^{(m+1)}_{\nu}({\rm s})=
\hat{R}_m \, \tilde{U}^{(1)}_{{\rm MA},\nu+m}({\rm s}),
\label{tdmp2N}
\eea
where $\tilde{U}^{(1)}_{{\rm MA},\nu}({\rm s})$ is given in Eq. (\ref{mainexpr}).

This approach allows to express the high order corrections in explicit form 
\be \tilde{\delta}^{(m+2)}_{\nu}({\rm s})=\frac{1}{(\nu+m)\pi(\pi^2+L_s^2)^{(\nu+m)/2}}
\Bigl\{\overline{\delta}^{(m+2)}_{\nu+m-1}({\rm s})\sin\bigl((\nu+m) g\bigr)
+ \hat{\delta}^{(m+2)}_{\nu+m-1}({\rm s}) g\cos\bigl((\nu+m) g\bigr)\Bigr\},
\label{tdm+2}
\ee
where $\overline{\delta}^{(m+2)}_{\nu}({\rm s})$ and $\hat{\delta}^{(m+2)}_{\nu}({\rm s})$ are
\bea
\z\overline{\delta}^{(2)}_{\nu}({\rm s})= b_1\,\Bigl[\hat{Z}_1(\nu)-G\Bigr],~~\hat{\delta}^{(2)}_{\nu}({\rm s})= b_1\,, \nonumber\\
\z\overline{\delta}^{(3)}_{\nu}({\rm s})= b_2+b_1^2\,\Bigl[\hat{Z}_2(\nu) -2G\hat{Z}_1(\nu)+G^2-g^2\Bigr],~~
\hat{\delta}^{(3)}_{\nu}({\rm s})= 2b_1^2\,\Bigl[\hat{Z}_1(\nu)-G\Bigr]\,, \nonumber\\
\z\overline{\delta}^{(4)}_{\nu}({\rm s})=\frac{b_3}{2} +3b_1b_2\,\Bigl[Z_1(\nu)- \frac{11}{6}-G\Bigl]  \nonumber\\
&&+ b_1^3\Bigl[\hat{Z}_3(\nu)-3G\hat{Z}_2(\nu)+3(G^2-g^2)\hat{Z}_1(\nu)-G(G^2-3g^2)\Bigr]\,, ~~\nonumber\\
\z\hat{\delta}^{(4)}_{\nu}({\rm s})=3b_1b_2 + b_1^3\Bigl[3\hat{Z}_2(\nu)-6G\hat{Z}_1(\nu)
  +(3G^2-g^2)\Bigr], \nonumber\\
\z\overline{\delta}^{(5)}_{\nu}({\rm s})=\frac{1}{3}\Bigl(b_4+5b_2^2\Bigr)+2b_1b_3\Bigl[Z_1(\nu)- \frac{13}{6}-G\Bigl] \nonumber \\
&&+6b_1^2b_2\Bigl[Z_2(\nu)- \frac{11}{3}Z_1(\nu)+\frac{38}{9}-2G\Bigl(Z_1(\nu)- \frac{11}{6}\Bigr)  +G^2-g^2\Bigl]\nonumber \\
\z+b_1^4\Biggl[\hat{Z}_4(\nu)-4G\hat{Z}_3(\nu)+6(G^2-g^2)\hat{Z}_2(\nu)
  -4G(G^2-3g^2)\hat{Z}_1(\nu)+G^4-6G^2g^2+g^4\Biggr], \nonumber\\
\z\hat{\delta}^{(5)}_{\nu}({\rm s})=2b_1b_3+12b_1^2b_2\Bigl[Z_1(\nu)- \frac{11}{6}-G\Bigl]\nonumber \\
&&+4b_1^4\Bigl[\hat{Z}_3(\nu)-3G\hat{Z}_2(\nu)
  +(3G^2-g^2)\hat{Z}_1(\nu)-G(G^2-g^2)\Bigr]
\label{ohdeltas}
\eea
and
\be
G({\rm s})=\frac{1}{2}\,\ln\left(\pi^2+L_s^2\right).
\label{G.def}
\ee

\subsection{The case $\nu=1$}
\label{nu1}

 For the case $\nu=1$ we get 
\be
\tilde{U}^{(i+1)}_{{\rm MA},\nu=1}({\rm s})=\tilde{U}^{(1)}_{{\rm MA},\nu=1}({\rm s}) + \sum_{m=1}^{i}\,
\tilde{\delta}^{(m+1)}_{\nu=1}({\rm s}),
\label{tdmp2N.1}
\ee
where LO gives the famous Shirkov-Solovtsov result \cite{ShS,Sh} 
   \be
   U^{(1)}_{{\rm MA}}({\rm s})=\tilde{U}^{(1)}_{{\rm MA},\nu=1}({\rm s})=\frac{g(s)}{\pi}=
\frac{1}{\pi}\arccos\left(\frac{L_s}{\sqrt{L_s^2+\pi^2}}\right)=\frac{1}{\pi}\left(\frac{\pi}{2}-\arctan\left(\frac{L_s}{\pi}\right)\right)
\label{Uv=1}
\ee
and the high order corrections sufficiently simplify
\be \tilde{\delta}^{(m+1)}_{\nu=1}({\rm s})=\frac{1}{m\pi(\pi^2+L_s^2)^{m/2}}
\Bigl\{\overline{\delta}^{(m+1)}_{m-1}({\rm s})\sin\bigl(m g\bigr)
+ \hat{\delta}^{(m+1)}_{m-1}({\rm s}) g\cos\bigl(m g\bigr)\Bigr\},
\label{tdm+2}
\ee
where $\overline{\delta}^{(m+1)}_{m-1}({\rm s})$ and $\hat{\delta}^{(m+1)}_{m-1}({\rm s})$ can be obtained from the corresponding values in Eq. (\ref{ohdeltas}) with $\nu=1$.
Using Eqs. (\ref{sin+cos}) and (\ref{sin+cos.1}) we get 
\bea
\z\overline{\delta}^{(2)}_{0}({\rm s})= -b_1\,\Bigl[1+G\Bigr],~~\hat{\delta}^{(2)}_{0}({\rm s})= b_1\,, \nonumber\\
\z\overline{\delta}^{(3)}_{1}({\rm s})= b_2+b_1^2\,\Bigl[G^2-g^2-1\Bigr],~~
\hat{\delta}^{(3)}_{1}({\rm s})= -2G\,b_1^2\,, \nonumber\\
\z\overline{\delta}^{(4)}_{2}({\rm s})=\frac{b_3}{2} -b_1b_2\,\Bigl[1+3G\Bigl]
+ \frac{b_1^3}{2}\Bigl[1+6G+3(G^2-g^2)-2G(G^2-3g^2)\Bigr]\,, ~~\nonumber\\
\z\hat{\delta}^{(4)}_{2}({\rm s})=3b_1b_2 + b_1^3\Bigl[3G^2-g^2-3G-3\Bigr], \nonumber\\
\z\overline{\delta}^{(5)}_{3}({\rm s})=\frac{1}{3}\Bigl(b_4+5b_2^2\Bigr)-\frac{2}{3}\,b_1b_3\Bigl[1+3G\Bigl]
+3b_1^2b_2\Bigl[2G^2-2g^2-1\Bigl]\nonumber \\
&&+b_1^4\Biggl[\frac{5}{3}+2G-4(G^2-g^2)
  -\frac{10}{3}G(G^2-3g^2)+G^4-6G^2g^2+g^4\Biggr], \nonumber\\
\z\hat{\delta}^{(5)}_{3}({\rm s})=2b_1b_3-12Gb_1^2b_2
+2b_1^4\Bigl[4G-1
  +\frac{5}{3}(3G^2-g^2)-2G(G^2-g^2)\Bigr]\,.
\label{ohdeltasA}
\eea

Another form of $\tilde{\delta}^{(i+1)}_{\nu=1}({\rm s})$ is given in Appendix A (see Eq. (\ref{nu1A})).\\

At the point $s=\Lambda^2$ the above results are simplified. They are
\be
\tilde{U}^{(i+1)}_{{\rm MA},\nu=1}({\rm{s}=\Lambda^2})=\tilde{U}^{(1)}_{{\rm MA},\nu=1}({\rm{s}=\Lambda^2}) + \sum_{m=1}^{i}\,
\tilde{\delta}^{(m+1)}_{\nu=1}({\rm s}=\Lambda^2),
\label{tdmp2N.1cri}
\ee
where LO gives
\be 
U^{(1)}_{{\rm MA},1}({\rm s}=\Lambda^2)= \frac{1}{2}
\label{Uv=1cri}
\ee
and the high order corrections are
\bea
&&\tilde{\delta}^{(2m+1)}_{\nu=1}({\rm s}=\Lambda^2)=\frac{(-1)^m}{4m(\pi^2+L_s^2)^{m}}
\hat{\delta}^{(2m+1)}_{2m-1}({\rm s}=\Lambda^2),~~\nonumber \\ 
&&\tilde{\delta}^{(2m+2)}_{\nu=1}({\rm s})=\frac{(-1)^m}{(2m+1)\pi(\pi^2+L_s^2)^{m+1/2}}\, \overline{\delta}^{(2m+2)}_{2m}({\rm s}=\Lambda^2)
\label{tdm+2cri}
\eea
where
$\overline{\delta}^{(2m+2)}_{2m}({\rm s}=\Lambda^2)$ and $\hat{\delta}^{(2m+1)}_{2m-1}({\rm s}=\Lambda^2)$ can be taken from Eq. (\ref{ohdeltasA})
with the following replacement:
\be
G(s=\Lambda^2)=\ln(\pi),~~g(s=\Lambda^2)=\frac{\pi}{2}\,.
\label{replycri}
\ee

\subsection{Discussions}
\label{Discu}

\begin{figure}[h!]
	\centering
	\includegraphics[width=0.68\textwidth]{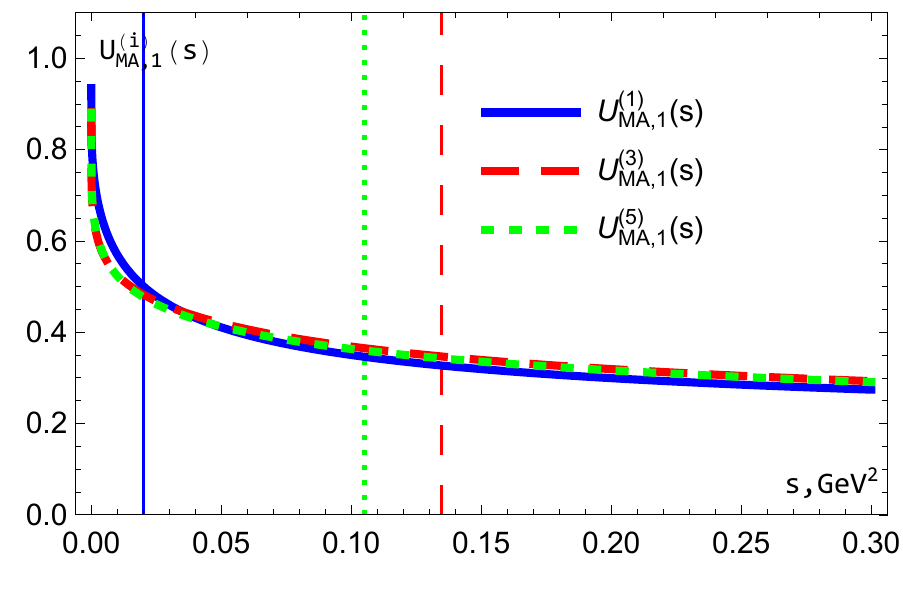}
	\caption{1,3 and 5 orders of $U_{{\rm MA},\nu=1}^{(i)}$ .The vertical lines indicate $(\Lambda^{f=3}_{i-1})^2$}
	\label{fig:u135}
\end{figure}
\begin{figure}[h!]
	\centering
	\includegraphics[width=0.68\textwidth]{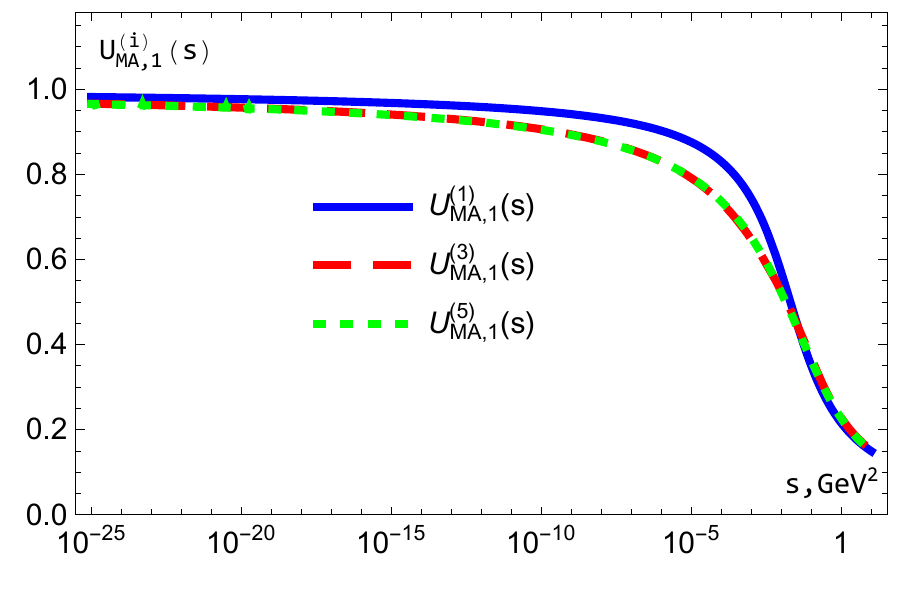}
	\caption{1,3 and 5 orders of $U_{{\rm MA},\nu=1}^{(i)}$ with logarithmic scale of $s$}
	\label{fig:logu135}
\end{figure}
\begin{figure}[h!]
	\centering
	\includegraphics[width=0.68\textwidth]{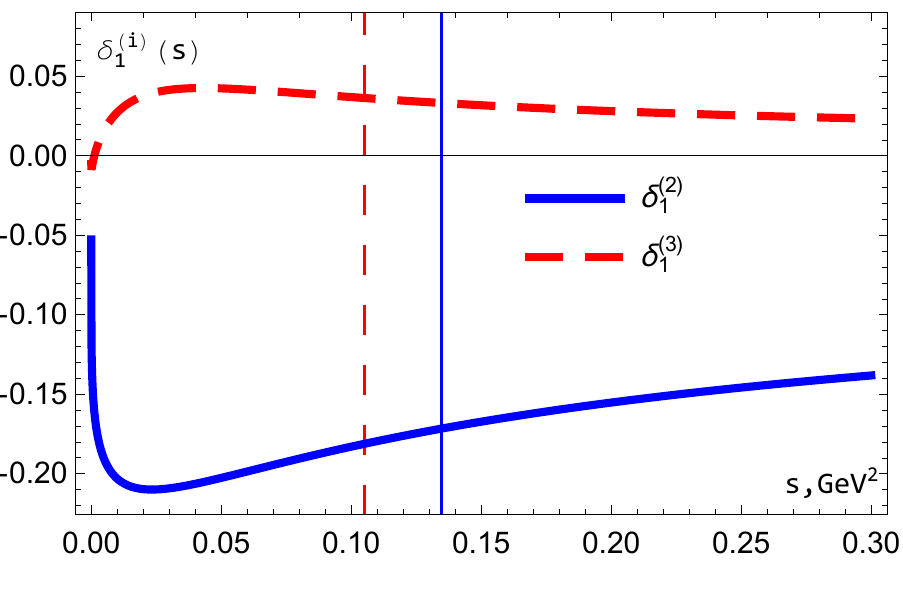}
	\caption{$\delta_{{\rm MA},\nu=1}^{(i)}$ with $i=2,3$.The vertical lines indicate $(\Lambda^{f=3}_{i-1})^2$}
	\label{fig:timecorr}
\end{figure}
\begin{figure}[h!]
	\centering
	\includegraphics[width=0.68\textwidth]{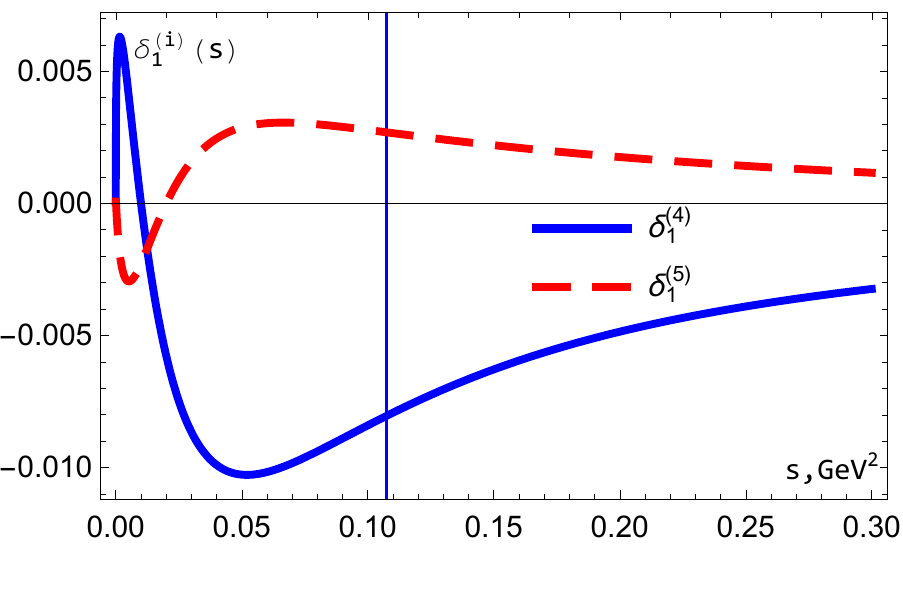}
	\caption{$\delta_{{\rm MA},\nu=1}^{(i)}$ with $i=4,5$. The vertical line indicates $(\Lambda^{f=3}_{3})^2=\Lambda^{f=3}_{4})^2$}
	\label{fig:timecorr45}
\end{figure}

\begin{figure}[h!]
	\centering
	\includegraphics[width=0.68\textwidth]{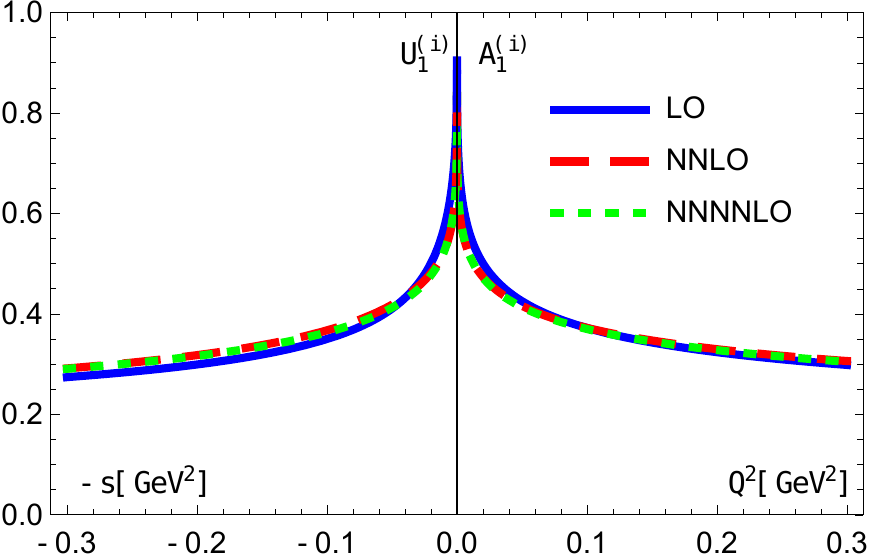}
	\caption{1,3 and 5 orders of $U_{{\rm MA},\nu=1}^{(i)}$ and $A_{{\rm MA},\nu=1}^{(i)}$}
	\label{fig:comparau}
\end{figure}
\begin{figure}[h!]
	\centering
	\includegraphics[width=0.68\textwidth]{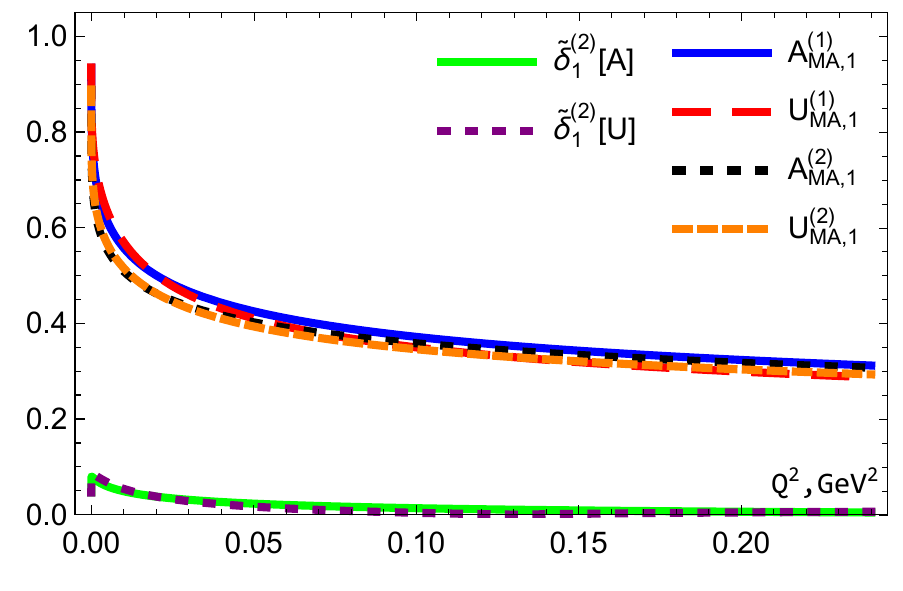}
	\caption{1 and 2 orders of $U_{{\rm MA},\nu=1}^{(i)}$, $A_{{\rm MA},\nu=1}^{(i)}$ and $\delta_{{\rm MA},\nu=1}^{(2)}$ in Euclidean and Minkowki spaces.}
	\label{fig:comparauN}
\end{figure}

\begin{figure}[h!]
	\centering
	\includegraphics[width=0.68\textwidth]{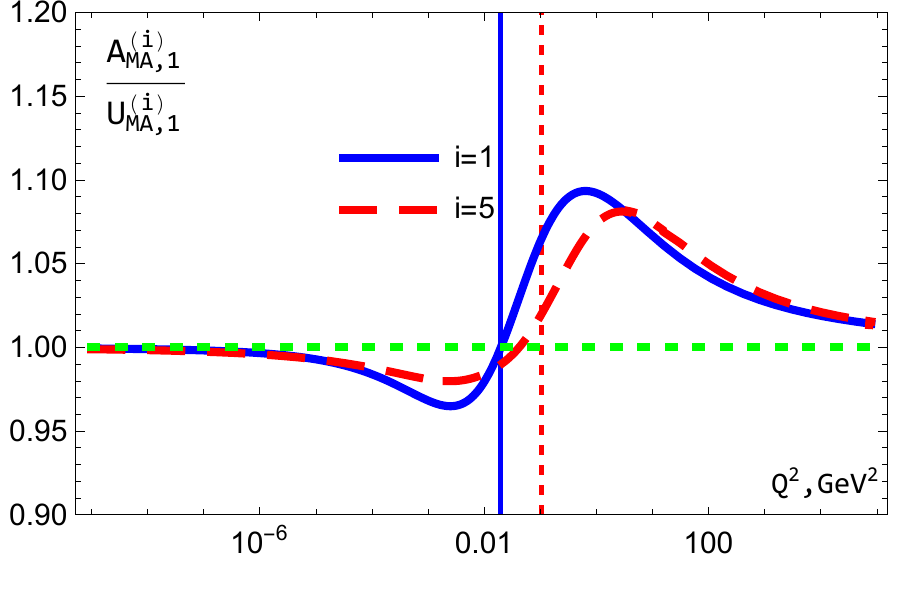}
	\caption{The relation $A^{(i)}_{\rm MA,\nu=1}/U^{(i)}_{\rm MA,\nu=1}$ for $i=1,5$.
        The vertical lines indicate $(\Lambda^{f=3}_{i-1})^2$.
        }
	\label{fig:aurel}
        \end{figure}

 This subsection provides graphical results of couplant construction.
Figs. \ref{fig:u135} and \ref{fig:logu135} show the results for $U^{(i)}_{\rm MA,\nu=1}(s)$ with $i=1,3,5$ in usual and logarithmic scales (the last one was chosen to stress the limit $U^{(i)}_{\rm MA,\nu=1}(s\to 0)\to1$).
From Figs. \ref{fig:timecorr} and \ref{fig:timecorr45} we can see the differences between $U^{(i)}_{\rm MA,\nu=1}(Q^2)$ with $i=1,..,5$,
which are rather small and have nonzero values around the
position $Q^2=\Lambda_i^2$.
In Figs. \ref{fig:u135}, \ref{fig:timecorr}, \ref{fig:timecorr45} and \ref{fig:aurel} the values of $(\Lambda_i^{f=3})^2$ $(i=0,2,4)$ are shown by vertical lines
  with color matching in each order. Note that Fig.\ref{fig:timecorr45} contains only one vertical line since $(\Lambda_4^{f=3})^2=(\Lambda_5^{f=3})^2$.

  So, Figs. \ref{fig:u135}-\ref{fig:timecorr45} point out that the difference between  $U^{(i+1)}_{\rm MA,\nu=1}(s)$ and $U^{(i)}_{\rm MA,\nu=1}(s)$  is essentially less then the couplants
  themselves. 
From Figs. \ref{fig:logu135}, \ref{fig:timecorr} and \ref{fig:timecorr45}
it is clear that for $s\to 0$ the asymptotic behavior of $U^{(1)}_{\rm MA,\nu=1}(s)$, 
$U^{(3)}_{\rm MA,\nu=1}(s)$  and $U^{(5)}_{\rm MA,\nu=1}(s)$ coincides (and is equal to behavior considered in (\ref{AQ=0})), i.e.
the differences $\delta^{(i)}_{\rm MA,\nu=1}(s\to 0)$  are negligible. 
Also  Figs. \ref{fig:timecorr} and \ref{fig:timecorr45} show the differences $\delta^{(i+1)}_{\rm MA,\nu=1}(s)$
$(i\geq 2)$ essentially less then $\delta^{(2)}_{\rm MA,\nu=1}(s)$.
We note that general form of the results is exactly the same as in the case of the MA couplants $A^{(i+1)}_{\rm MA,\nu,i}(Q^2)$, which have been studied earlier in
\cite{Kotikov:2022sos}.
         Indeed, the similarity is shown in Figs. \ref{fig:comparau} and \ref{fig:comparauN}. In Fig.  \ref{fig:comparau}  the results for
  $U^{(i)}_{\rm MA,\nu=1}(s)$ and  $A^{(i)}_{\rm MA,\nu=1}(Q^2)$ ($i=1,3,5$) are shown in 
         the so-called mirror form, which is in accordance with the similar one presented
         earlier in \cite{Bakulev:2006ex}. Fig. \ref{fig:comparauN} contains
  $U^{(1)}_{\rm MA,\nu=1}(s)$, $A^{(1)}_{\rm MA,\nu=1}(Q^2)$,
  $U^{(2)}_{\rm MA,\nu=1}(s)$ and $A^{(2)}_{\rm MA,\nu=1}(Q^2)$ which are very close to each others but have different limit values when $Q^2\to 0$. Moreover, the differences $\delta^{(2)}_{\rm MA,\nu=1}(Q^2)$ in the cases
$U^{(2)}_{\rm MA,\nu=1}(s)$ and $A^{(2)}_{\rm MA,\nu=1}(Q^2)$ are almost the same although correction of the spacelike couplant decreases more rapidly. The direct relation between $A^{(i)}_{\rm MA,\nu=1}(Q^2)$ and $U^{(i)}_{\rm MA,\nu=1}(Q^2)$ gives an interesting picture (see Fig. \ref{fig:aurel}). Obviously we have $\frac{A^{(i)}_{\rm MA,\nu=1}(Q^2=0)}{U^{(i)}_{\rm MA,\nu=1}(Q^2=0)}=1$ for any order and the second similar point
\be
\frac{A^{(i)}_{\rm MA,\nu=1}(Q^2=(\Lambda^{f=3}_{i-1})^2)}{U^{(i)}_{\rm MA,\nu=1}(Q^2=(\Lambda^{f=3}_{i-1})^2)}=1
\label{AUrel}
\ee
for $i=1$. Higher order corrections break the identity (\ref{AUrel}), shifting the second point from $(\Lambda^{f=3}_i)^2$.
    As we can see in Fig. \ref{fig:aurel}, the shift is quite small.
As can be seen from Fig. \ref{fig:aurel}, the ratio (\ref{AUrel}) asymptotically approaches 1 when Q$^ 2$ $\to \infty$.

Thus, we can conclude that contrary to the case of the usual couplant,
the $1/L$-expansion of the MA couplant is very good approximation at any $Q^2(s)$ values.
Moreover, the differences between $U^{(i+1)}_{\rm MA,\nu=1}(s)$ and $U^{(i)}_{\rm MA,\nu=1}(s)$ become smaller with the increase of order.
So, the expansions of  $U^{(i+1)}_{\rm MA,\nu=1}(s)$ $i\geq 1$ through the $U^{(1)}_{\rm MA,\nu=1}(s)$ done in Refs. \cite{BMS1,Bakulev:2006ex,Bakulev:2010gm} are
very good approximations.

\section{Integral representations for minimal analytic coupling }
\label{IntegralRe}

As it was abovementioned in Introduction, the MA couplants $A^{(1)}_{\rm MA,\nu}(Q^2)$ and $U^{(1)}_{\rm MA,\nu}(s)$ are constructed as follows: 
the LO
spectral function is taken directly from perturbation theory
but the MA couplants $A^{(1)}_{\rm MA,\nu}(Q^2)$ and $U^{(1)}_{\rm MA,\nu}(s)$ themselves were obtained using the correct integration contours.
Thus, at LO, the MA couplants $A^{(1)}_{\rm MA,\nu}(Q^2)$ and $U^{(1)}_{\rm MA,\nu}(s)$  obey Eqs. (\ref{disp_MA_LO}) and (\ref{disp_MAt_LO})  presented in Introduction.

To check Eqs. (\ref{tdm+2}) and (\ref{ohdeltas}) we compare them with an integral form 

\be 
U^{(i)}_{1}({\rm s})= \frac{1}{\pi}\int\limits^{\infty}_{s}\frac{d\sigma}{\sigma}r^{(i)}_{{\rm pt}}(\sigma).
\label{integral.nu=1}
\ee

For LO, we can take the integral form from \cite{Bakulev:2006ex}
\be 
U^{(1)}_{\nu}({\rm s})= \frac{1}{\pi}\int\limits^{\infty}_{s}\frac{d\sigma}{\sigma}r^{(1)}_{\nu}(\sigma),
\label{integral.nu}
\ee
where
\be
r^{(1)}_\nu({\rm s})=\frac{\sin[\nu\,g(s)]
}{\pi(\pi^2+ L^2_s)^{(\nu-1)/2}} = \nu \, U^{(1)}_{\nu+1}({\rm s})
  \, ,
\label{rLOnu}
\ee

In (\ref{mainexpr}) only the case $\nu\ge0$ is considered, it means that the integral (\ref{integral.nu}) converges to zero at the upper limit.  We would like to note, that
dispersion integral (\ref{integral.nu=1}) does not converge
for some $\nu$ and, in this case, we will introduce constant, which corresponds to the upper limit of the integral.
In general situation, it is better to replace the integral (\ref{integral.nu=1}) by the one
\be 
U^{(1)}_{\nu}({\rm s})=  \frac{1}{\pi}\int\limits^{\infty}_{s}\frac{d\sigma}{\sigma}r^{(1)}_{\nu}(\sigma)-U^{(1)}_{\nu}(\infty),
\label{integral.nu.n}
\ee
where
\be
U^{(1)}_{\nu}(\infty) = \left\{
\begin{array}{c}
	0 ~~\mbox{when}~~ \nu >0, \\
	1 ~~\mbox{when}~~ \nu =0, \\
	\infty ~~\mbox{when}~~ \nu <0. 
\end{array}
\right.
\label{Unuinfty}
\ee
We see that the expression (\ref{integral.nu}) diverges for $\nu <0$ and requires additional constant for $\nu =0$. Therefore Eq. (\ref{mainexpr}) is applicable only
when $\nu >0$.
Further in this paper we will only consider the region $\nu>0$.

Using our approach to obtain high-order terms from LO  (\ref{integral.nu}), we can extend the LO integral  (\ref{integral.nu}) to the one
\be 
\tilde{U}^{(i)}_{\nu}({\rm s})= \frac{1}{\pi}\int\limits^{\infty}_{s}\frac{d\sigma}{\sigma}r^{(i)}_{\nu}(\sigma),
\label{int.nu}
\ee
where obviously
\be
r^{(i)}_\nu({\rm s})= \nu \, \tilde{U}^{(i)}_{\nu+1}({\rm s})
  \, .
\label{rnu}
\ee

The spectral function $r^{(i)}_1({\rm s})$ has the form
\be
r^{(i)}_1({\rm s})=r^{(1)}_1({\rm s}) +  \sum_{m=1}^{i}\, 
\delta^{(m+1)}_{1}({\rm s})\,
\label{rnu=1}
\ee
where
\be
r^{(1)}_1({\rm s})=   U^{(1)}_{2}({\rm s}),~~
\delta^{(m+1)}_{1}({\rm s})=(m+1)\, \tilde{\delta}^{(m+1)}_{\nu=2}({\rm s}).
\label{deltar}
\ee
In the explicit form:
\bea
\z r^{(i)}_1({\rm s})= \frac{\sin(g)}{\pi(\pi^2+L_s^2)^{1/2}}= \frac{1}{\pi^2+L_s^2}, \nonumber\\
\z \tilde{\delta}^{(m+1)}_{1}({\rm s})=\frac{1}{\pi(\pi^2+L_s^2)^{(m+1)/2}}
\Bigl\{\overline{\delta}^{(m+1)}_{m}({\rm s})\sin\bigl((m+1) g\bigr)
+ \hat{\delta}^{(m+1)}_{m}({\rm s}) g\cos\bigl((m+1) g\bigr)\Bigr\},
\label{Rnu.1}
\eea
where $\overline{\delta}^{(m+1)}_{m}({\rm s})$ and $\hat{\delta}^{(m+1)}_{m}({\rm s})$ can be obtained from the results in (\ref{ohdeltas}) with $\nu=2$. They are
\bea
\z\overline{\delta}^{(2)}_{1}({\rm s})= -G\,b_1,~~\hat{\delta}^{(2)}_{1}({\rm s})= b_1\,, \nonumber\\
\z\overline{\delta}^{(3)}_{2}({\rm s})= b_2+b_1^2\,\Bigl[G^2-g^2-G-1\Bigr],~~
\hat{\delta}^{(3)}_{2}({\rm s})= b_1^2 \,\Bigl[1-2G\Bigr]\,, \nonumber\\
\z\overline{\delta}^{(4)}_{3}({\rm s})=\frac{b_3}{2} -3G\,b_1b_2
+ \frac{b_1^3}{2}\Bigl[4G-1+5(G^2-g^2)-2G(G^2-3g^2)\Bigr]\,, ~~\nonumber\\
\z\hat{\delta}^{(4)}_{3}({\rm s})=3b_1b_2 + b_1^3\Bigl[3G^2-g^2-5G-2\Bigr], \nonumber\\
\z\overline{\delta}^{(5)}_{4}({\rm s})=\frac{1}{3}\Bigl(b_4+5b_2^2\Bigr)-2\,b_1b_3\Bigl[\frac{1}{12}+G\Bigl]
+3b_1^2b_2\Bigl[2G^2-2g^2-G-1\Bigl]\nonumber \\
&&+b_1^4\Biggl[\frac{7}{6}+4G-\frac{3}{2}(G^2-g^2)
  -\frac{13}{3}G(G^2-3g^2)+G^4-6G^2g^2+g^4\Biggr], \nonumber\\
\z\hat{\delta}^{(5)}_{4}({\rm s})=2b_1b_3+ 3 b_1^2b_2\Bigl[1-4G\Bigr]
+b_1^4\Bigl[3G-4
  +\frac{13}{3}(3G^2-g^2)-4G(G^2-g^2)\Bigr]\,.
\label{ohdeltas.1}
\eea

Using the results  (\ref{sin+cos}) and (\ref{sin+cos.1}) for $\cos(ng)$ and $\sin(ng)$ $(n\leq 4)$, we see that
with the results \cite{NeSi,Nesterenko:2017wpb}  (see also Section 6 in  \cite{Kotikov:2022sos})
%
give more compact results for $r^{(i)}_1({\rm s})$.
We think that Eqs. (\ref{Rnu.1}) and (\ref{ohdeltas.1})
give apparently very compact results for $r^{(i)}_1({\rm s})$.

Note that the results (\ref{int.nu}) for $\tilde{U}^{(i)}_{\nu}({\rm s})$ are exactly the same as the results in Eq. (\ref{tdmp2N}) done in the form of
  trigonometric factions.
 However the results (\ref{int.nu}) should be very handy in case of non-minimal versions of analytic couplants
 (see Refs. \cite{Cvetic:2006mk,Cvetic:2006gc,Cvetic:2010di}).

\section{$H \to b\overline{b}$ decay}
\label{HbbDecay}

In Ref. \cite{Kotikov:2022sos} we used the polarized Bjorken sum rule \cite{Chen:2006tw} as an example for the application of
the MA couplant $A_{\rm MA}(Q^2)$,
which is a popular object of study in the framework of analytic QCD (see \cite{Pasechnik:2008th,Ayala:2017uzx,Ayala:2018ulm,Kotikov:2012eq}). Here we consider the
decay of the Higgs boson into a bottom-antibottom pair, which is also a popular application of the MA couplant $U_{\rm MA}( Q^2)$ (see, e.g., \cite{Bakulev:2006ex}
and reviews in Ref. \cite{Bakulev:2008td}).

The Higgs-boson decay into a bottom-antibottom pair can be expressed in QCD by means
of the correlator
\be
\Pi(Q^2)=(4\pi)^2i\int \, dx e^{iqx} <0|T[J_b^S(x)J_b^S(0)]0>
\label{Pi}
\ee
of two quark scalar (S) currents in terms of the discontinuity of its imaginary part, i.e.,
$R_S(s)={\bf Im}\Pi(-s-i\ep)/(2\pi s)$, so that the width reads
\be
\Gamma(H \to b\overline{b})=\frac{G_F}{4\sqrt{2}\pi} M_H m^2_b(M^2_H) R_s(s=M^2_H)\,.
\label{GHbb}
\ee

Direct multi-loop calculations were
performed in the Euclidean (spacelike) domain for the corresponding Adler
function $D_S$ (see Refs. \cite{Broadhurst:2000yc,Chetyrkin:1996sr,Baikov:2005rw,Chetyrkin:1997wm}).
Hence, we write ($D_s \to \tilde{D}_s$ and $R_s \to \tilde{R}_s$ because the additional factor
$m_b^2$)
\be
\tilde{D}(Q^2)=3m_b^2(Q^2)\Bigl[1+\sum_{n\geq 1} d_n \, a^n_s(Q^2)\Bigr],
\label{GHbb}
\ee
where for $f=5$ the coefficients $d_n$ are
\be
d_1=2.96,~~d_2=11.44,~~d_2=50.17,~~d_4=260.24,~~
\label{dk}
\ee

Taking the imagine part, one has
\be
\tilde{R}_s(s)=3m_b^2(s)\Bigl[1+\sum_{n\geq 1} r_n \, a^n(s)\Bigr],
\label{Hbb}
\ee
and  for $f=5$ \cite{Baikov:2005rw,Kataev:1993be}
\be
r_1=2.96,~~r_2=7.93,~~r_3=5.93,~~r_4=-61.84,~~
\label{ri}
\ee

Here $\overline{m}_b^2(Q^2)$ has the form (see Appendix B):
\be
\overline{m}_b^2(Q^2)=\hat{m}_b^2 a_s^d(Q^2) \,\left[ 1+ \sum_{k=1}^{k=4} \, \overline{e}_k \, a_s^k(Q^2)\right]\,,
\label{omb}
\ee
where
\be
\overline{e}_k=\frac{\tilde{e}_k}{k(\beta_0)^k}
\label{oek}
\ee
and $\tilde{e}_k$ are done in Eq. (\ref{tei}). For $f=5$ we have
\be
\overline{e}_1=1.23,~~\overline{e}_2=1.20,~~\overline{e}_3=0.55,~~\overline{e}_4=0.54\,.~~
\label{oekf5}
\ee

The normalization constant $\hat{m}_b$ cab be obtained as (see, e.g., \cite{Bakulev:2008td})
\be
\hat{m}_b=\overline{m}_b(Q^2=m_b^2) a_s^{-d/2}(m_b^2) \,{\left[ 1+ \sum_{k=1}^{k=4} \, \overline{e}_k \, a_s^k(Q^2)\right]}^{-1/2}=10.814~ \mbox{GeV}^2\,,
\label{hmb}
\ee
since $\overline{m}_b(Q^2=m_b^2)=m_b=4.18$ GeV.

So, we have
\be
\tilde{R}_s(s)=\tilde{R}^{(m=5)}_s(s),~~\tilde{R}^{(m+1)}_s(s)=
3\hat{m}_b^2\, a_s^d(s) \, \left[1+\sum_{k=0}^m \overline{r}_k a_s^k(s))\right],
\label{Hbb.1}
\ee
where
\be
\overline{r}_k=r_k +\overline{e}_k + \sum_{l=1}^{k-1} r_l \, \overline{e}_{k-l}\,.
\label{ork}
\ee
 For $f=5$ we have
\be
\overline{r}_1=4.18,~~\overline{r}_2=12.76,~~\overline{r}_3=19.76,~~\overline{r}_4=-42.25\,.~~
\label{orkf5}
\ee

We can express all results through derivatives $\tilde{a}_{d+k}$ (see Appendix B):
\be
\tilde{R}_s(s)=\tilde{R}^{(m=5)}_s(s),~~\tilde{R}^{(m+1)}_s(s)=
3\hat{m}_b^2 \left[\tilde{a}_d +\sum_{k=0}^m \tilde{r}_k \tilde{a}_{d+k}\right],
\label{Hbb.2}
\ee
where
\be
\tilde{r}_k=\overline{r}_k +\tilde{k}_k(d) + \sum_{l=1}^{k-1} \overline{r}_l \, \tilde{k}_{k-l}(d+l)\,,
\label{trk}
\ee
where $\tilde{k}_i(\nu)$ are given in Appendix C.

For $d=24/23$ and $f=5$, we have
\be
\tilde{r}_1=4.17,~~\tilde{r}_2=9.86,~~\tilde{r}_3=1.29,~~\tilde{r}_4=-71.21\,.
\label{trkf5}
\ee

Performing the same analysis for the Adler function we have
\be
\tilde{D}_s=\tilde{D}^{(m=5)}_s,~~\tilde{D}^{(m+1)}_s=
3\hat{m}_b^2 a_s^d(Q^2) \, \left[1+\sum_{k=0}^m \overline{d}_k a_s^k(Q^2))\right],
\label{tDs}
\ee
where
\be
\overline{d}_k=r_k +\overline{e}_k + \sum_{l=1}^{k-1} d_l \, \overline{e}_{k-l}\,.
\label{odk}
\ee
 For $f=5$ we have
\be
\overline{d}_1=4.18,~~\overline{d}_2=16.27,~~\overline{d}_3=68.30,~~\overline{r}_4=337.66\,.~~
\label{odkf5}
\ee

We express all results through derivatives $\tilde{a}_{d+k}$:
\be
\tilde{D}_s=\tilde{D}^{(m=5)}_s,~~\tilde{D}^{(m+1)}_s=
3\hat{m}_b^2 \left[\tilde{a}_d(Q^2) +\sum_{k=0}^m\tilde{d}_k \tilde{a}_{d+k}(Q^2)\right],
\label{tDs.2}
\ee
where
\be
\tilde{d}_k=\overline{d}_k +\tilde{k}_k(d) + \sum_{l=1}^{k-1} \overline{d}_l \, \tilde{k}_{k-l}(d+l)\,.
\label{tdk}
\ee

For $f=5$ and $d=24/23$, we have
\be
\tilde{d}_1=4.17,~~\tilde{d}_2=13.37,~~\tilde{d}_3=43.90,~~\tilde{d}_4=178.18 \,.~~
\label{trkf5}
\ee

As it was discussed earlier in \cite{Bakulev:2006ex} in FAPT there are the following representation for $\tilde{R}_s$
\be
\tilde{R}_s(s)=\tilde{R}^{(m=5)}_s(s),~~\tilde{R}^{(m+1)}_s(s)=
3\hat{m}_b^2 \left[\tilde{U}^{(m+1)}_{d}({\rm s}) +\sum_{k=0}^m \tilde{d}_k \, \tilde{U}^{(m+1)}_{d+k}(s)\right],
\label{tRsFART}
\ee

\begin{figure}[h!]
	\centering
	\includegraphics[width=0.68\textwidth]{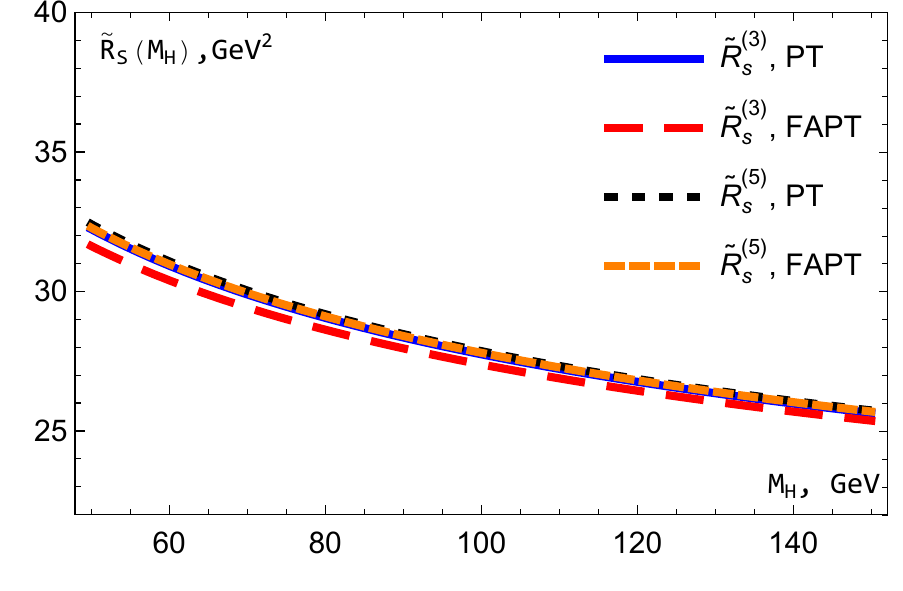}
	\caption{The results for $\tilde{R}^{(m+1)}_s(s)$ with $m=2$ and 4 in the framework of the usual PT and FAPT.}
	\label{fig:comparau1}
\end{figure}

The results for $\tilde{R}^{(m+1)}_s(s)$ are shown in Fig. \ref{fig:comparau1}. We see that the FAPT results (\ref{tRsFART})
are lower than those (\ref{Hbb.2}) based on the conventional PT.
  This is in full agreement with arguments given in \cite{Bakulev:2008td}.
  But the difference becomes less notable as the PT order increases.  Indeed, for N$^3$LO the difference is very small, which
  proves the assumption about the possibility of using $\tilde{R}^{(m+1)}_s(s)$ expression
  for $\tilde{D}^{(m+1)}_s(Q^2)$ with $A^{(i)}_{\rm MA}(Q^2) \to U^{(i)}_{\rm MA}(s)$,
  which was done in Ref. \cite{Bakulev:2006ex}.
  
The results for $\Gamma^{(m)}(H \to b\overline{b})$
 in the N$^{m}$LO approximation using $\tilde{R}^{(m+1)}_s(s)$ from Eqs. (\ref{Hbb.1}) and (\ref{Hbb.2}) are exactly same and have the following form:
\bea
&&\Gamma^{(0)}=1.76~\mbox{MeV},~\Gamma^{(1)}~=2.27~\mbox{MeV},~~\Gamma^{(2)}=2.37~\mbox{MeV},~~\nonumber \\
&&\Gamma^{(3)}=2.38~\mbox{MeV},~~\Gamma^{(4)}=2.38~\mbox{MeV}\,.
\label{tR.nu}
\eea
The corresponding
results for $\Gamma^{(m)}(H \to b\overline{b})$
with $\tilde{R}^{(m+1)}_s(s)$ form Eq. (\ref{tRsFART}) are very similar to ones in (\ref{tR.nu}).
They are:
\bea
&&\Gamma^{(0)}=1.74~\mbox{MeV},~\Gamma^{(1)}~=2.23~\mbox{MeV},~~\Gamma^{(2)}=2.34~\mbox{MeV},~~\nonumber \\
&&\Gamma^{(3)}=2.37~\mbox{MeV},~~\Gamma^{(4)}=2.38~\mbox{MeV}\,.
\label{tR.nu.1}
\eea

So, we see a good agreement between the results obtained in FAPT and in the framework of the usual PT.
  
It is clearly seen that the results of FAPT are very also close to the results \cite{Wang:2013bla}
  obtained in the framework of the now very popular Principle of Maximum Conformality \cite{Brodsky:2011ta} (for the recent review, see \cite{Shen:2022nyr}).
  Indeed, our results are within the band obtained by varying the renormalization scale.

The Standard Model expectation is \cite{LHCHiggsCrossSectionWorkingGroup:2016ypw}
  \be
  \Gamma^{SM}_{H\to b\overline{b}}(M_H=125.1 \mbox{GeV}) =2.38 \mbox{MeV}\,.
  \label{GaSM}
\ee
The ratios of the measured events yield to the Standard Model expectations are $1.01 \pm 0.12 (\mbox{stat}.) + 0.16 - 0.15 (\mbox{syst}.)$ \cite{ATLAS:2018kot}
  in ATLAS Collaboration
and $1.04 \pm 0.14 (\mbox{stat}.) \pm 0.14 (\mbox{syst}.)$ \cite{CMS:2018nsn} in SMC Collaboration (see also \cite{Tsukerman:2020qwz}).

  Thus, our results obtained in both approaches, in the standard perturbation theory and in analytical QCD, are in
  good agreement both with the  Standard Model expectations \cite{LHCHiggsCrossSectionWorkingGroup:2016ypw} and with the experimental data
  \cite{ATLAS:2018kot,CMS:2018nsn}.

\section{Conclusions}
\label{Conclu}

In this paper
we have used $1/L$-expansions of the $\nu$-derivatives of the strong couplant $a_s$ expressed \cite{Kotikov:2022sos} as
combinations of operators $\hat{R}_m$ (\ref{hR_i}) applied to the LO couplant $a_s^{(1)}$.
Applying the same operators to the $\nu$-derivatives of the LO
MA couplant $U_{\rm MA}^{(1)}$,  we obtained two different
representations (see Eqs. (\ref{tdm+2}) and (\ref{int.nu}))
for the $\nu$-derivatives of the MA couplants, i.e. $\tilde{U}_{\rm MA, \nu}^{(i)}$ introduced for timelike processes, in each
$i$-order of perturbation theory:
 one form contains a combinations of trigonometric functions,
 and the other  is based on dispersion integrals containing
the  $i$-order spectral function.
All results are presented up to the 5th order of perturbation theory, where the corresponding coefficients of the QCD $\beta$-function are well known
(see \cite{Baikov:2016tgj,Herzog:2017ohr}).

As in the case of $\tilde{A}_{\rm MA, \nu}^{(i)}$ \cite{Kotikov:2022sos} applied in the Euclidean space,
high-order corrections for $\tilde{U}_{\rm MA, \nu}^{(i)}$ are negligible in the $s \to 0$ and $s \to \infty$ \textcolor{red}{limits}
and are nonzero in  the vicinity of
the point $s =\Lambda^2$. Thus,  in fact, there
are actually only small corrections to the LO MA couplant $U_{\rm MA,\nu}^{(1)}(s)$.
In particular, this proves the possibility of expansions of high-order couplants $U_{\rm MA,\nu}^{(i)}(s)$ via the LO couplants $U_{\rm MA,\nu}^{(1)}(s)$,
which was done in Ref. \cite{Bakulev:2010gm}.

As an example, we examined the Higgs boson decay into a $b\overline{b}$ pair and obtained results are in good agreement with  the
Standard Model expectations \cite{LHCHiggsCrossSectionWorkingGroup:2016ypw} and with the experimental data  \cite{ATLAS:2018kot,CMS:2018nsn}.
Moreover, our results also in good agreement with
studies based on the Principle Maximum
Conformality \cite{Brodsky:2011ta}. 

As a next step, we plan to include $1/L$-expansions for other MA couplants
(see Refs. \cite{Bakulev:2006ex,Bakulev:2010gm,Ayala:2018ifo,Mikhailov:2021znq}), as well as for non-minimal analytic couplants
(following Refs. \cite{Cvetic:2006mk,Cvetic:2006gc,Cvetic:2010di,CPCCAGC,3dAQCD}).
In the case of non-minimal analytic couplants, one can use the integral representations (\ref{integral.nu}) and (\ref{int.nu}) with non-perturbative spectral functions.

  \section{Acknowledgments}
  We are grateful to Andrey Kataev for discussions.
  We also thank the anonymous Referee, whose comments greatly improved the quality of the paper.
  This work was supported in part by the Foundation for the Advancement of Theoretical
Physics and Mathematics “BASIS”.

\appendix
\def\theequation{A\arabic{equation}}
\setcounter{equation}{0}

\section{Another form for $U^{(i+1)}_1({\rm s})$}
\label{Anotherform}

Using the results in Eq. (\ref{Ls}) and transformation rules for $\sin(ng)$ and $\cos(ng)$, we have
\be
\sin(ng)=\frac{S(ng)}{(\pi^2+L_s^2)^{n/2}},~~\cos(ng)=\frac{C(ng)}{(\pi^2+L_s^2)^{n/2}},
\label{sin+cos}
\ee
where
\bea
&&S(g)=\pi,~~C(g)=L_s,~~S(2g)=2\pi L_s,~~C(2g)=L_s^2-\pi^2,~~S(3g)=\pi(3L_s^2-\pi^2),~\nonumber \\
&&~C(3g)=L_s(L_s^2-3\pi^2),~~S(4g)=4\pi L_s(L_s^2-\pi^2),~~C(4g)=L_s^4-6\pi^2L_s^2+\pi^4,~\nonumber \\
&&S(5g)=\pi(5L_s^4-10\pi^2L_s^2+\pi^4),~~C(5g)=L_s(L_s^4-10\pi^2L_s^2+5\pi^4)\,.
\label{sin+cos.1}
\eea


Using Eqs. (\ref{sin+cos}), (\ref{sin+cos.1}) and (\ref{ohdeltasA}), the results for $\tilde{\delta}^{(m+1)}_{\nu=1}({\rm s})$ in (\ref{tdm+2}) can be rewritten in the
following form
\bea
\z \tilde{\delta}^{(2)}_{\nu=1}({\rm s}) = \frac{b_1}{(\pi^2+L_s^2)}\Bigl(\frac{g}{\pi}L_s-\bigl(1+G\bigr)\Bigr), \nonumber\\
\z \tilde{\delta}^{(3)}_{\nu=1}({\rm s}) = \frac{1}{(\pi^2+L_s^2)^2}\Bigl[b_2 L_s-b_1^2\Bigl(\frac{gG}{\pi}(L_s^2-\pi^2)+L_s\bigl(1+g^2-G^2\bigr)\Bigr)\Bigr],\nonumber\\
\z \tilde{\delta}^{(4)}_{\nu=1}({\rm s}) = \frac{1}{(\pi^2+L_s^2)^3}\Bigl\{b_1b_2\Bigl(\frac{gL_s}{\pi}(L^2_s-3\pi^2)-\Bigl(L^2_s-\frac{1}{3}\pi^2\Bigr)\bigl(1+3G\bigr)\Bigr)
\nonumber\\
&&+\frac{b_3}{6}(3L^2_s-\pi^2)+\frac{b_1^3}{6}\Bigl((3L_s^2-\pi^2)\Bigl[1+6G-3g^2+3G^2+6g^2G-2G^3\Bigr]\nonumber\\
&&-\frac{2gL_s}{\pi}(L^2_s-3\pi^2)\Bigl[3+3G+g^2-3G^2\Bigr]\Bigr)\Bigr\},\nonumber\\
  \z \tilde{\delta}^{(5)}_{\nu=1}({\rm s}) = \frac{1}{(\pi^2+L_s^2)^4}\Bigl\{3b_1^2b_2\Bigl(L_s(\pi^2-L_s^2)\Bigl[1+2g^2-2G^2\Bigr]-\frac{gG}{\pi}(L_s^4-6L_s^2\pi^2+\pi^4)\Bigr)\nonumber\\
&&+2b_1b_3\Bigl(\frac{L_s}{3}(\pi^2-L_s^2)\Bigl[1+3G\Bigr]+\frac{g}{4\pi}(L_s^4-6L_s^2\pi^2+\pi^4)\Bigr)-\frac{5b_2^2+b_4}{3}L_s(\pi^2-L^2_s)\nonumber\\
&&+b_1^4\Bigl(L_s(\pi^2-L^2_s)\Bigl[-\frac{5}{3} -2G-4g^2+4G^2-10g^2G+\frac{10}{3}G^3-g^4+6g^2G^2-G^4\Bigr]\nonumber\\
&&+\frac{g}{2\pi}(L_s^4-6L_s^2\pi^2+\pi^4)\Bigl[4G-1-\frac{5}{3}g^2+5G^2+2g^2G-2G^3\Bigr]\Bigr)\Bigr\},
\label{nu1A}
\eea
which is similar to the results for the spectral function $r^{(i)}_1({\rm s})$
done in ref.\cite{NeSi,Nesterenko:2017wpb}  (see also Section 6 in  \cite{Kotikov:2022sos}).

\def\theequation{B\arabic{equation}}
\setcounter{equation}{0}

\section{$\overline{m}_b^2(Q^2)$}
\label{mbQ2}

Here we present evaluation of $\overline{m}_b^2(Q^2)$, which has the form
\be
\overline{m}_b^2(Q^2)=\overline{m}_b^2(Q^2) \, \exp\left[2\int^{\oar(Q^2)}_{\oar(Q^2_0)} \frac{\gamma_m(a)}{\beta_(a)}\right]\,,~~ \oar(Q^2)=\frac{\alpha_s(Q^2)}{4\pi}\,,
\label{omb}
\ee
where
\bea
&&\gamma_m(a)=-\sum_{k=0} \gamma_{k}a^{k+1} =- \gamma_{0}a \bigl(1+\sum_{k=1} \delta_{k}a^{k}\bigr),~~ \delta_{k}=\frac{\gamma_{k}}{\gamma_{0}}\,, \nonumber \\
&&\beta(a)=-\sum_{k=0} \beta_{k}a^{k+2} =- \beta_{0}a^2 \bigl(1+\sum_{k=1} c_{k}a^{k}\bigr),~~ c_{k}=\frac{\beta_{k}}{\beta_0}\,.
\label{gb}
\eea

Evaluating the integral in (\ref{omb}) we have the following results (see, e.g., also Refs. \cite{Bakulev:2006ex,Chetyrkin:1997dh})
\be
\overline{m}_b^2(Q^2)=\overline{m}_b^2(Q^2) \, \frac{\oar^d(Q^2)}{\oar^d(Q_0^2)} \, \frac{T(\oar(Q^2))}{T(\oar(Q^2_0))}\,,
\label{omb1}
\ee
where
\be
d=\frac{2\gamma_0}{\beta_0},~~T(\oar)=\exp[\sum_{k=1}^{k=4} \frac{e_k}{k}\,\oar^k ]
\label{dT}
\ee
and
\bea
&&e_1=d\Delta_1,~~e_2=d(\Delta_2-c_1\Delta_1),~~e_3=d(\Delta_3-c_1\Delta_2-\tilde{c}_2\Delta_1),\nonumber \\
&&e_4=d(\Delta_4-c_1\Delta_3-\tilde{c}_2\Delta_2-\tilde{c}_3\Delta_1)\,,
\label{di}
\eea
with
\be
\Delta_i=\delta_i-c_i,~~ \tilde{c}_2=c_2-c_1^2,~~\tilde{c}_3=c_3-2c_1c_2+c_1^3
\label{Deltai}
\ee

The result for $T(\oar)$ can be rewritten as
\be
T(\oar)=1+ \sum_{k=1}^{k=4} \, \frac{\tilde{e}_k}{k} \, \oar^k\,,
\label{T}
\ee
where
\bea
&&\tilde{e}_1=e_1,~~\tilde{e}_2=e_2+e_1^2,~~\tilde{e}_3=e_3+\frac{3}{2}e_1e_2+\frac{1}{2}e_1^3,\nonumber \\
&&\tilde{e}_3=e_4+e_2^2+\frac{4}{3}e_1e_3+\frac{1}{2}e_1^3+ e^2_1e_2+\frac{1}{6}e_1^4 ,
\label{tei}
\eea

\def\theequation{C\arabic{equation}}
\setcounter{equation}{0}

\section{Relations between $a_s^{\nu}$ and  $\tilde{a}_{\nu}$}
\label{astoTildeas}

Considering Ref. \cite{GCAK} we have
\be
a_s^{\nu}= \tilde{a}_{\nu} + \sum_{m\geq 1} \, \tilde{k}_m(\nu) \, \tilde{a}_{\nu+m}\,,
\label{atota}
\ee
where
\bea
\z \tilde{k}_1(\nu) = -\nu b_1 \tilde{B}_1(\nu), \nonumber \\
\z \tilde{k}_2(\nu) = \nu(\nu+1) \left(- b_2 \tilde{B}_2(\nu) + \frac{b_1^2}{2} \tilde{B}_{1,1}(\nu)\right),\nonumber \\
\z \tilde{k}_3(\nu) = \frac{\nu(\nu+1)(\nu+2)}{2} \left(- b_3 \tilde{B}_3(\nu) + b_1b_2 \tilde{B}_{1,2}(\nu)- \frac{b_1^3}{3} \tilde{B}_{1,1,1}(\nu)\right),\nonumber \\
\z \tilde{k}_4(\nu) = \frac{\nu(\nu+1)(\nu+2)(\nu+3)}{6} \Biggl(- b_4 \tilde{B}_4(\nu) + b_2^2 \tilde{B}_{2,2}(\nu)+ \frac{b_1b_3}{2} \tilde{B}_{1,3}(\nu)\nonumber \\
&&- \frac{b_1^2b_2}{2} \tilde{B}_{1,1,2}(\nu)+ \frac{b_1^4}{4} \tilde{B}_{1,1,1,1}(\nu)\Biggr),
\label{tkm}
\eea
where
\bea
\z  \tilde{B}_1(\nu)= \tilde{Z}_1(\nu)-1,~~ \tilde{B}_2(\nu)= \frac{\nu-1}{2(\nu+1)},~~ \tilde{B}_{1,1}(\nu)= \tilde{Z}_2(\nu)-2\tilde{Z}_1(\nu+1)+1, \nonumber \\
\z \tilde{B}_3(\nu)= \frac{1}{6}-\frac{1}{(\nu+1)(\nu+2)},~~ \tilde{B}_{1,2}(\nu)= \frac{\nu-1}{6(\nu+1)} \,
\Bigl(6\tilde{Z}_1(\nu+1)-1+ \frac{4}{\nu+2}\Bigr), \nonumber \\
\z \tilde{B}_{1,1,1}(\nu)= \tilde{Z}_3(\nu)-3\tilde{Z}_2(\nu+1)+3\tilde{Z}_1(\nu+2)-1, \nonumber \\
\z  \tilde{B}_4(\nu)= \frac{1}{12}-\frac{2}{(\nu+1)(\nu+2)(\nu+3)},~~\tilde{B}_{2,2}(\nu)= \frac{13}{12}-\frac{1}{\nu+1}-\frac{1}{\nu+2}-\frac{1}{\nu+3}, \nonumber \\
\z \tilde{B}_{1,3}(\nu)= \left(1-\frac{6}{(\nu+1)(\nu+2)}\right) \tilde{Z}_1(\nu+3) +  \frac{1}{6}+\frac{4}{\nu+1}-\frac{5}{\nu+2}-\frac{2}{\nu+3}, \nonumber \\
\z \tilde{B}_{1,1,2}(\nu)=  \frac{3(\nu-1)}{2(\nu+1)} \, \tilde{Z}_2(\nu+2)-\left(1-\frac{6}{(\nu+1)(\nu+2)}\right) \tilde{Z}_1(\nu+3) \nonumber \\
&&+  \frac{8}{3}
-\frac{2}{\nu+1}+\frac{1}{\nu+2}-\frac{8}{\nu+3}, \nonumber \\
\z \tilde{B}_{1,1,1,1}(\nu)= \tilde{Z}_4(\nu)-4\tilde{Z}_3(\nu+1)+6\tilde{Z}_2(\nu+2)-4\tilde{Z}_1(\nu+3)+1
\label{tBm}
\eea
and
\bea
\z \tilde{Z}_1(\nu)=S_1(\nu),~~\tilde{Z}_2(\nu)=S^2_1(\nu)+S_2(\nu), \nonumber \\
\z \tilde{Z}_3(\nu)=S^3_1(\nu)+3S_2(\nu)S_1(\nu)+2S_3(\nu), \nonumber \\
\z \tilde{Z}_4(\nu)=S^4_1(\nu)+6S_2(\nu)S^2_1(\nu)+3S_2^2(\nu)+8S_3(\nu)S_1(\nu)+6S_4(\nu)\,.
\label{tZm}
\eea

For arbitrary $\nu$ values, $S_i(\nu)$ are expressed through Polygamma-functions as
\bea
\z S_1(\nu)=\Psi(\nu+1)-\Psi(1),~~\Psi(\nu)\equiv \frac{d}{d\nu} \ln \Gamma(\nu),~~ \Psi^{(i)}(\nu)\equiv \frac{d^i}{(d\nu)^{i}} \Psi(\nu), \nonumber \\
\z S_{i+1}(\nu) \equiv \frac{(-1)^{i}}{i!} \Bigl(\Psi^{(i)}(\nu+1)-\Psi^{(i)}(1)\Bigr).
\label{Sinu}
\eea

In the case of integer $\nu=n$,
\be
S_i(n)=\sum_{m=1}^{n}\, \frac{1}{m^i}\,.
\label{Sin}
\ee

\end{document}